\documentclass[hyper]{JHEP3}
\usepackage[dvips]{graphicx}
\usepackage{epsfig,subfigure}
\usepackage{latexsym,amsmath,amssymb}

\title{Area spectrum of Schwarzschild black hole inspired by noncommutative geometry}

\author{Shao-Wen Wei,
        Yu-Xiao Liu\footnote{Corresponding author.},
        Zhen-Hua Zhao,
        Chun-E Fu\\
   Institute of Theoretical Physics, Lanzhou University,
           Lanzhou 730000, P. R. China \\
  E-mail: \email{weishaow06@lzu.cn}
          \email{liuyx@lzu.edu.cn}
          \email{zhaozhh09@lzu.cn}
          \email{fuche08@lzu.cn}}

\date{\today}

\abstract{It is known that, in the noncommutative Schwarzschild black hole spacetime, the point-like object is replaced by the smeared object, whose mass density is described by a Gaussian distribution of minimal width $\sqrt{\theta}$ with $\theta$ the noncommutative parameter. The elimination of the
point-like structures makes it quite different from the conventional Schwarzschild black hole. In this paper, we mainly investigate the area spectrum and entropy spectrum for
the noncommutative Schwarzschild black hole with $0\leq \theta\leq \left(\frac{M}{1.90412}\right)^{2}$. By the use of the new physical interpretation of
the quasinormal modes of black holes presented by Maggiore, we obtain the quantized area spectrum and entropy spectrum with the modified Hod's and Kunstatter's methods, respectively. The results show that: (1) The area spectrum and entropy spectrum are discrete. (2) The spectrum spacings are dependent on the parameter $\frac{M}{\sqrt{\theta}}$. (3) The spacing of the area spectrum of the noncommutative Schwarzschild black hole is smaller than that of the conventional one. So does the spacing of the entropy spectrum. (4) The spectra from the two methods are consistent with each other.}

\begin{document}

\section{Introduction}
\label{Introduction}

Recently, motivated by string theory arguments \cite{Snyder1947pr},
noncommutative geometry has been studied extensively. An important
application of the noncommutative geometry is the black hole spacetime.
In the noncommutative black hole spacetime, the point-like structure is eliminated \cite{Smailagic2003jpa} and the point-like object is replaced by the smeared object.

The first noncommutative black hole solution, known as the noncommutative Schwarzschild black hole, was presented by Nicolini, Smailagic and Spallucci \cite{Nicolini2005plb} five years ago. It was found that there exists no singularity of scalar curvature $R$ at the origin, which is different from the conventional Schwarzschild black hole for the temperature diverges and the scalar curvature
becomes arbitrarily large \cite{Nicolini2005jpa}. Further, one can show that there exist no scalar singularities in this noncommutative black hole background according to the classification of the singularities given by Cai and Wang \cite{Cai10plb}. Thermodynamic properties of the noncommutative black hole were studied in
\cite{Giri2007ijmpla,Banerjee2008prd,Nozari2008cqg,Kim2008jhep,Vakili2009ijmpd,
Buric2008epjc,Huang2009plb,Park2009prd}. All these results showed interesting
features for the noncommutative black hole at the small radius,
while they coincide with the conventional black hole at the large
radius. The evaporation of the noncommutative black hole was studied
in \cite{Garcia2006prd},
where it was shown that the final remnant of an extremal black hole
is a thermodynamically stable object, which is different from that
of the conventional Schwarzschild black hole for no remnant exists at
the end of the evaporation. It was also observed
\cite{Banerjee2008prd} that, in the regime
$\frac{M}{\sqrt{\theta}}\gg 1$, the noncommutative entropy/area
law will recover the standard Bekenstein-Hawking area law, i.e.,
$S=\frac{A}{4}$.

Another important difference between the
noncommutative Schwarzschild black hole and the conventional one is
that the noncommutative one may have two horizons, one degenerate
horizon or no horizon for different values of parameters. So, in
some sense, the noncommutative black hole behaves like those black
holes with two horizons. The similarity of the thermodynamics
between it and the Reissner-Nordstr\"om (RN) black hole was studied
and the noncommutative parameter $\theta$ is identified as the
charge of the black hole with a simple relation \cite{Kim2008jhep}.
Other noncommutative black hole solutions were found and their
thermodynamics were investigated in
\cite{Rizzojhep2006,Grezia2009}.

Motivated by the earlier work, we would like to study the area and entropy spectra for the noncommutative Schwarzschild black hole and we want to known whether the noncommutative parameter $\theta$ has any effect one the area and entropy spectra. With the new physical interpretation of
the quasinormal modes of black holes presented by Maggiore
\cite{Maggioreprl2008}, we obtain the quantized area spectrum and
entropy spectrum with the modified Hod's and Kunstatter's
methods \cite{Hodprl1998,Kunstatterprl2003}, respectively. The results show that the noncommutative parameter $\theta$ indeed has impact on the area spectrum and entropy spectrum. In the small $\frac{M}{\sqrt{\theta}}$ limit, the spacings of the area and entropy spectra have significant difference between the nonmcommutative and conventional Schwarzschild black holes. And when $\frac{M}{\sqrt{\theta}}\gg 1$, the area and entropy spectra, respectively, are consistent with each other for the two black holes.

The paper is organized as follows. In Sec. \ref{noncommutative}, we
introduce the noncommutative black hole and examine its
thermodynamic quantities in detail. With the new interpretation of the quasinormal modes, we calculate the area spectrum and entropy spectrum for the
noncommutative black hole in Sec. \ref{entropy}. Finally, the paper ends with a brief summary.

\section{Review of the noncommutative Schwarzschild black hole solution}
\label{noncommutative}

\begin{figure*}
\centerline{\subfigure[]{\label{gammafig}
\includegraphics[width=6.5cm,height=5cm]{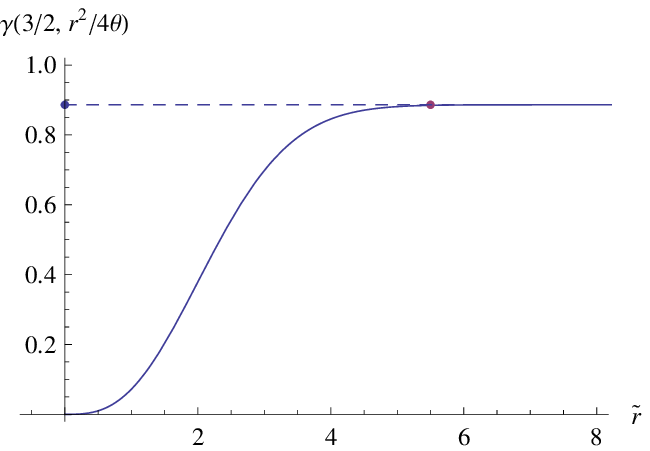}}
\subfigure[]{\label{metricfunction}
\includegraphics[width=6.5cm,height=5cm]{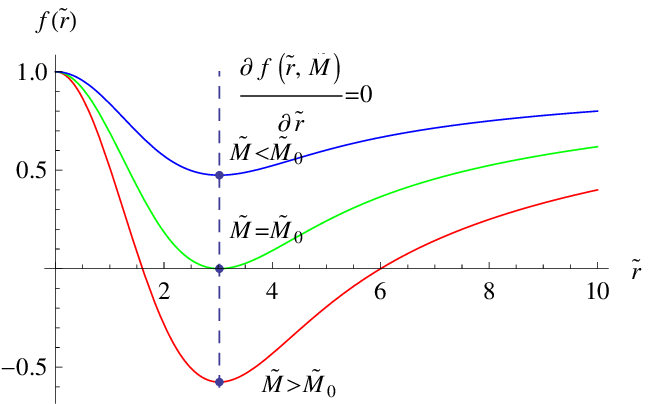}}}
\caption{(a) The lower incomplete Gamma function vs
$\tilde{r}=\frac{r}{\sqrt{\theta}}$. When $\frac{r}{\sqrt{\theta}}$ is greater
than 5.5, $\gamma(\frac{3}{2},\frac{\tilde{r}^{2}}{4})$ will close
to it's maximum $\frac{\sqrt{\pi}}{2}$. (b) The metric function $f(\tilde{r})$ vs
$\tilde{r}$ with different values of the redefined mass $\tilde{M}$.}
\end{figure*}

In this section, we would like to give a brief review on the
noncommutative black hole. An important property of the
noncommutative black hole spacetime is that it eliminates the point-like
structures \cite{Smailagic2003jpa}, and which is
replaced by a smeared object and the effect of smearing is
mathematically implemented by replacing the position Dirac-delta
function with a Gaussian distribution of minimal width
$\sqrt{\theta}$ everywhere. Inspired by this idea, the mass density
of a static, spherically symmetric, smeared, particle-like
gravitational source is thought in the form \cite{Nicolini2005plb}
\begin{eqnarray}
 \rho_{\theta} = \frac{M}{(4\pi\theta)^{\frac{3}{2}}} \exp \left(
    -\frac{r^2}{4\theta} \right).
\end{eqnarray}
The total mass $M$ is diffused throughout the region
of linear size $\sqrt{\theta}$. Then the mass involved in a sphere
with radius $r$ is
\begin{eqnarray}
 m(r)&=&\int_{0}^{r} 4\pi r^{2}\rho_{\theta}dr\nonumber\\
     &=&\frac{2M}{\sqrt{\pi}}\gamma(\frac{3}{2},\frac{r^{2}}{4\theta}),
\end{eqnarray}
where $\gamma(\frac{3}{2},\frac{r^{2}}{4\theta})$ is the lower
incomplete Gamma function defined by
\begin{eqnarray}
  \gamma (\frac{3}{2}, \frac{r^{2}}{4\theta})
     \equiv  \int_{0}^{\frac{r^{2}}{4\theta}} t^{\frac{1}{2}} e^{-t} dt.
\end{eqnarray}
The energy-momentum tensor $T^{\mu}_{\;\;\;\nu}$ describing a
static, spherically symmetric noncommutative black hole spacetime is \cite{Nicolini2005plb}
\begin{eqnarray}
  T^{\mu}_{\;\;\;\nu}=\text{diag}\bigg(-\rho_{\theta}, -\rho_{\theta},
         -\rho_{\theta}-\frac{1}{2}r\partial_{r}\rho_{\theta},
         -\rho_{\theta}-\frac{1}{2}r\partial_{r}\rho_{\theta}\bigg),\label{tensor}
\end{eqnarray}
which is found to satisfy the conservation condition
$T^{\mu\nu};\nu=0$. The metric of the noncommutative Schwarzschild black
hole is taken as
\begin{eqnarray}
  ds^{2} = -f(r) dt^{2} + f(r)^{-1} dr^{2}
         + r^{2} (d\vartheta^{2}
             +\sin^{2}\vartheta d\varphi^{2}).\label{metric}
\end{eqnarray}
Solving the Einstein equations with this metric and the
energy-momentum tensor (\ref{tensor}), we could obtain the explicit
form of the metric function $f(r)$, which is given by
\cite{Nicolini2005plb}
\begin{eqnarray}
  f(r)=1-\frac{2m(r)}{r} = 1 - \frac{4M}{r\sqrt{\pi}} \gamma\bigg( \frac32,
    \frac{r^2}{4\theta} \bigg).\label{methricf}
\end{eqnarray}
Note that this black hole spacetime is closely dependent on the
noncommutative parameter $\theta$. However, there should  be a
natural restriction that the metric (\ref{metric}) should recover
the conventional Schwarzschild black hole as
$\theta\rightarrow 0$.
For such purpose, we plot the lower incomplete Gamma function in
Fig. \ref{gammafig}. We could see that when $r/\sqrt{\theta}$ is
greater than 5.5, the value of Gamma function will be closer to
$\frac{\sqrt{\pi}}{2}$. Thus the metric function becomes
$f(r)\approx 1-\frac{2M}{r}$ for small $\theta$, which is just the conventional
Schwarzschild case.

%%%%%%%%%%%%%%%%%%%%%%%%%%%%%%%%%%%%%%%%%%%%%%%%%%%%%%%%%%%%%%%%%%%%%%%%%%%%%%
%%%%%%%%%%%%%%%%%%%%%%%%%%%%%%%%%%%%%%%%%%%%%%%%%%%%%%%%%%%%%%%%%%%%%%%%%%%%%%%%

From (\ref{methricf}), it is obvious that the metric function
contains the noncommutative parameter $\theta$, however we can
hide it by introducing the redefined mass and radial
coordinate:
\begin{eqnarray}
  \tilde{M}=\frac{M}{\sqrt{\theta}},\quad \tilde{r}=\frac{r}{\sqrt{\theta}}.
\end{eqnarray}
Then, the metric function becomes
\begin{eqnarray}
  f(\tilde{r})= 1 - \frac{4\tilde{M}}{\tilde{r}\sqrt{\pi}}
  \gamma\bigg( \frac32,\frac{\tilde{r}^2}{4}\bigg),
\end{eqnarray}
where the noncommutative parameter $\theta$ has been successfully
hidden. The behavior of $f(\tilde{r})$ is plotted against the radial
coordinate $\tilde{r}$ with different values of $\tilde{M}$ in
Fig. \ref{metricfunction}. The horizons are determined by
$f(\tilde{r})=0$. So, the horizons are located at the zero points of
$f(\tilde{r})=0$, which are shown in Fig. \ref{metricfunction}. Form
the figure, we could explicitly see that there are two horizons for
$\tilde{M}>\tilde{M}_{0}$, while one degenerate horizon for
$\tilde{M}=\tilde{M}_{0}$ and no horizon for
$\tilde{M}<\tilde{M}_{0}$. Thus, $\tilde{M}_{0}$ can be regarded as an extremal
mass for it splits the whole region into the non-extremal black hole
and the naked singularity regions. Assume the black hole event horizon is existed, then it is given
by
\begin{eqnarray}
  \tilde{r}_{h}=\frac{4\tilde{M}}{\sqrt{\pi}}
                \gamma\bigg(\frac{3}{2},
                \frac{\tilde{r}_{h}^{2}}{4}\bigg).
\end{eqnarray}
This is an iterative equation and has no analytical solution.
However, one could obtain an approximate solution with an iteration
method, i.e.,
\begin{eqnarray}
  r_{h}= 2M\left[1-e^{-\frac{M^{2}}{\theta}}
          \left( \frac{2M}{\sqrt{\pi\theta} }
              +  \mathcal{O}\big(\frac{\sqrt{\theta}}{M}\big)
          \right) \right].
\end{eqnarray}
Now, we would like to determine the values of the extremal mass
$\tilde{M}_{0}$ and the corresponding radius $\tilde{r}_{0}$. From
Fig. \ref{metricfunction}, we could see that there exists one
minimum value of $f(\tilde{r})$ for each mass $\tilde{M}$. When
$\tilde{M}>\tilde{M}_{0}$, the minimum is negative, while it
is zero and is positive for $\tilde{M}=\tilde{M}_{0}$ and
$\tilde{M}<\tilde{M}_{0}$, respectively. So, the extremal point must
lie on the line determined by
\begin{eqnarray}
  \frac{\partial f(\tilde{r})}{\partial \tilde{r}}=0.
  \label{condition1}
\end{eqnarray}
Another condition to determine the extremal point is
\begin{eqnarray}
  f(\tilde{r})=0.\label{condition2}
\end{eqnarray}
By solving these two equations (\ref{condition1}) and
(\ref{condition2}), we can determine $\tilde{M}_{0}$ and
$\tilde{r}_{0}$ uniquely. Substituting the metric function
$f(\tilde{r})$ into (\ref{condition1}), we derive
\begin{eqnarray}
  \tilde{r}^{3}e^{-\frac{\tilde{r}^{2}}{4}}
      =4\gamma
      \bigg(\frac{3}{2},\frac{\tilde{r}^{2}}{4}\bigg).\label{condition11}
\end{eqnarray}
Interestingly, this equation does not contain the mass parameter
$\tilde{M}$. So, the line determined by (\ref{condition1}) is
essentially a vertical line in Fig. \ref{metricfunction}. It is not
hard to understand that the root of Eq. (\ref{condition11}) is just
the value of $\tilde{r}_{0}$ and we obtain $\tilde{r}_{0}=3.02244$
numerically. Substituting $\tilde{r}_{0}$ into (\ref{condition2}),
we get the extremal mass $\tilde{M}_{0}=1.90412$. Or, the extremal
point can also be expressed as $r_{0}=3.02244\sqrt{\theta}$ and
$M_{0}=1.90412\sqrt{\theta}$. This result is exactly consistent with
that of \cite{Kim2008jhep}. On the other hand, we can obtain the range $\theta$ for a black hole,
\begin{eqnarray}
 &&0\leq\theta\leq\left(\frac{M}{1.90412}\right)^{2}.\label{Range}
\end{eqnarray}

Following, we would like to examine the thermodynamics quantities
for this noncommutative black hole solution. The Hawking temperature
$T$ defined by $T=\frac{\partial_{r}f(r)}{4\pi}\mid_{r_{h}}$ is read
\begin{eqnarray}
  T&=&\frac{1}{4\pi r_{h}}\bigg( 1-
          \frac{r_{h}^{3}}
             {4\theta^{3/2}\gamma(3/2,r_{h}^{2}/4\theta)} e^{-r_{h}^{2}/4\theta} \bigg)\nonumber\\
   &=&\frac{1}{8\pi M}-\frac{M}{2\sqrt{\pi^{3}}\theta}e^{-M^{2}/\theta}
    \left( \frac{M}{\sqrt{\theta} }
              +  \mathcal{O}\big(\frac{\sqrt{\theta}}{M}\big)
          \right).\label{temperature}
\end{eqnarray}
When $\frac{M}{\sqrt{\theta}}\gg 1$, it will return to the conventional
Schwarzschild case and give $T_{\text{Sch}}=\frac{1}{8\pi M}$. It was shown that the Benkenstein-Hawking entropy/area law is held
for the noncommutative black hole \cite{Banerjee2008prd}. Thus, the
entropy is
\begin{eqnarray}
  S&=&\frac{A}{4}=\pi r_{h}^{2}\nonumber\\
               &=& 4\pi M^{2}-16\sqrt{\pi} M^2 e^{-\frac{M^{2}}{\theta}}
          \left( \frac{M}{\sqrt{\theta} }
              +  \mathcal{O}\big(\frac{\sqrt{\theta}}{M}\big)
          \right).
\end{eqnarray}
Heat capacity is a key quantity to measure the thermal stability of
a black hole. Generally, a black hole with positive heat capacity
can be stable existed in a heat bath, while a negative one will be
all evaporated when a perturbation appears. Form
(\ref{temperature}), we can obtain the heat capacity:
\begin{eqnarray}
  C_{\theta}=\bigg(\frac{dT}{dM}\bigg)^{-1}=-\frac{1}{8\pi M^{2}}
        +\frac{M^{2}-\theta}{\sqrt{\pi^{3}}\theta^{2}}e^{-\frac{M^{2}}{\theta}}
  \left( \frac{M}{\sqrt{\theta} }
              +  \mathcal{O}\big(\frac{\sqrt{\theta}}{M}\big)
          \right).
\end{eqnarray}
For the conventional Schwarzschild black hole, the heat capacity
$C_{\text{Sch}}=-\frac{1}{8\pi M^{2}}$. The negative heat capacity $C_{\text{Sch}}$
implies an unstable black hole. However, for the noncommutative
black hole, the capacity $C_{\theta}$ is positive for $M\in(1.90412\sqrt{\theta},2.3735\sqrt{\theta})$. So, the
noncommutative black hole can stably exist in the range. It is also clear that, for $\frac{M}{\sqrt{\theta}}\gg 1$, the heat capacity $C_{\theta}$ will recover the conventional one.

Here, we have checked the thermodynamics quantities for the
noncommutative black hole. It is shown that, different from the conventional one, the noncommutative black hole can stably exist in a heat bath in some range. It is also
worth to point out that, for $\frac{M}{\sqrt{\theta}}\gg 1$, each thermodynamics
quantity of the noncommutative black hole is identical to that of
the conventional black hole.

\section{Area and entropy spectra of noncommutative Schwarzschild black hole}
\label{entropy}

The quantization of the black hole horizon area and entropy is an
old but very interesting topic. Hod was one of the first to consider
this problem ten years ago. He combined the perturbations of black
holes with the principles of quantum mechanics and statistical
physics in order to derive the quantum of the black hole area
spectrum. With this idea, he obtained the area spectrum
$A=4l_{p}^{2}\ln 3\cdot n$ \cite{Hodprl1998}.

On the other hand, Bekenstein first pointed out that the black hole
horizon area is an adiabatic invariant \cite{Bekenstein1997} and the
spacing of the area spectrum obtained from this viewpoint is $\Delta
A=8\pi l_{p}^{2}$. Moreover, given a system with energy $E$ and vibrational frequency
$\omega(E)$, the ratio $\frac{E}{\omega(E)}$ is a nature adiabatic
invariant \cite{Kunstatterprl2003}. And using the Bohr-Sommerfeld
quantization, Kunstatter \cite{Kunstatterprl2003} derived an equally
spaced entropy spectrum for the $D$-dimensional Schwarzschild black
hole. Subsequently, Hod's and Kunstatter's methods
rejuvenated the interest on the study of the quantization of black
hole area and entropy and the methods had been extended to other
black holes
\cite{Dreyerprl2003,Vagenasjhep2008,Medvedcqg2008,Kothawalaprd2008,Abdalla2003ijmpla,
Wei2009,Wei2010a,Myung2010,Kwon2010,li2009plb,Setare2010}.

Very recently, Maggiore presented a new physical interpretation for
the quasinormal modes of black holes \cite{Maggioreprl2008}. He
suggested that the proper frequency of the equivalent harmonic
oscillator, which is interpreted as the quasinormal mode frequency
$\omega(E)$, should be of the form:
\begin{eqnarray}
 \omega(E)=\sqrt{|\omega_{R}|^{2}+|\omega_{I}|^{2}}.
\end{eqnarray}
The form of the proper frequency for the quasinormal modes was first
presented in \cite{Wang2004prd}. Note that, for the case of
long-lived quasinormal modes $(\omega_{I}\rightarrow 0)$, we have
$\omega(E)=|\omega_{R}|$, approximately. However, for the case of
highly excited quasinormal modes $(|\omega_{I}|\gg |\omega_{R}|)$,
the natural selection should be $\omega(E)=|\omega_{I}|$. With this
new physical interpretation of the quasinormal modes, the area
spectrum of the Kerr black hole was obtained by Vagenas
\cite{Vagenasjhep2008} with the modified Hod's and
Kunstatter's methods, respectively. The spacing of the area spectrum
calculated with the modified Hod's method is equally spaced,
while it is non-equidistant and depends on the angular momentum
parameter $J$ employing the Kunstatter's method. The two methods
give different spacings of the area spectrum. At the same time, it
was Medved \cite{Medvedcqg2008} who pointed out the difference of
these results. He argued that the Kunstatter's method is only
effective for the non-extremal Kerr black hole. The reason is that
the quantum number $n$ appearing in the Bohr-Sommerfeld quantization
condition is actually a measure of the areal deviation from
extremality for the black hole. Thus, the calculation of the
Kunstatter's method is restricted to the case $M^{2}\gg J$. In the
spirit of this idea, the two methods give the same area spectrum,
which is equally spaced and angular momentum parameter
$J$-independent area spectrum.

Other charged or rotating black holes were also studied with the two
methods \cite{Myung2010}. Following the Kunstatter's
method, the equally spaced area spectra were obtained for the
non-extremal black holes. For the stringy charged
Garfinkle-Horowitz-Strominger black hole, we
showed that the area spectrum and entropy spectrum are both equally
spaced and independent of the charge $q$ \cite{Wei2010a}. For other
non-Einstein gravity theories, the entropy spectra were found to be
equally spaced, while the area spectra were non-equidistant (the
detail can be found in
\cite{Kothawalaprd2008,Wei2009,Banerjee,Setare2010}).

Note that the works discussed above are all for conventional black
holes. Now we would like to investigate the area spectrum and
entropy spectrum for the noncommutative black hole. Different from
the conventional black hole, we want to know that whether the
noncommutative parameter $\theta$ has any effect on the area
spectrum and entropy spectrum of the noncommutative black hole.

With these questions, we start our calculation. First, we will study
the area spectrum and entropy spectrum for the noncommutative black
hole by using the modified Hod's method. The asymptotic
quasinormal frequency for the noncommutative black hole has been
obtained in \cite{Giri2007ijmpla}:
\begin{eqnarray}
 \omega=T\ln 3+i 2\pi T\bigg(k+\frac{1}{2}\bigg).
 \label{quasinormal}
\end{eqnarray}
For the temperature is closely dependent on the noncommutative
parameter $\theta$, so the quasinormal frequency $\omega$ also
depends on the parameter $\theta$. The change in the parameters of
the noncommutative black hole is determined by
\begin{eqnarray}
 \Delta M=\hbar \Delta \omega,\label{masschange}
\end{eqnarray}
where $\Delta\omega$ can be obtained from Eq. (\ref{quasinormal}).
Considering the transition $k-1\rightarrow k$ for the noncommutative
black hole, we obtain
\begin{eqnarray}
 \Delta\omega\approx |\omega_{I}|_{k}-|\omega_{I}|_{k-1}=2\pi T,\quad (k\gg 1).
 \label{vibrational}
\end{eqnarray}
Generally, the change in the black hole mass will create a change in
the black hole area:
\begin{eqnarray}
 \Delta A=  32\pi M
   \left[1
        +\frac{4M^{2}-6\theta}{\sqrt{\pi}\theta}
          e^{-\frac{M^{2}}{\theta}}
          \left( \frac{M}{\sqrt{\theta} }
              +  \mathcal{O}\big(\frac{\sqrt{\theta}}{M}\big)
          \right)
   \right]   \Delta M.\label{areachange}
\end{eqnarray}
Recalling the expression of the temperature in (\ref{temperature}), Eq. (\ref{areachange}) can be be rewritten as
\begin{eqnarray}
 \Delta A=\frac{4}{T}\Delta M
   +e^{-\frac{M^{2}}{\theta}}\mathcal{O}\big(\frac{\sqrt{\theta}}{M}\big). \label{areachange1}
\end{eqnarray}
Substituting (\ref{masschange}) into (\ref{areachange1}), we obtain
the spacing of the area spectrum
\begin{eqnarray}
 \Delta A=8 \pi \hbar
 +e^{-\frac{M^{2}}{\theta}}\mathcal{O}\big(\frac{\sqrt{\theta}}{M}\big).\label{areaspings}
\end{eqnarray}
Neglecting the high order of $\frac{\sqrt{\theta}}{M}$, we can obtain a $\theta$- independent spacing of the area spectrum. The area spectrum for the noncommutative black hole can be assumed in the form:
\begin{eqnarray}
 A_{n}=8 \pi \hbar\cdot n
   +e^{-\frac{M^{2}}{\theta}}\mathcal{O}\big(\frac{\sqrt{\theta}}{M}\big). \label{areaspectrum}
\end{eqnarray}
Correspondingly, the quantized entropy spectrum is obtained through
the entropy/area law:
\begin{eqnarray}
 S_{n}=2 \pi \hbar\cdot n
     +e^{-\frac{M^{2}}{\theta}}\mathcal{O}\big(\frac{\sqrt{\theta}}{M}\big). \label{entropyspectrum}
\end{eqnarray}
A couple of comments are in order here. First, the area and entropy spectra are dependent on the parameter $\theta$. Second, neglecting the high order of $\frac{\sqrt{\theta}}{M}$ (i.e., the case that far from the extremal black hole), we get equally spaced area and entropy spectra, which is in full agreement with that of the Schwarzschild black hole given by
Maggiore \cite{Maggioreprl2008}.

Next, we will study the area spectrum and entropy spectrum for the
noncommutative black hole by employing the Kunstatter's method. The
adiabatic invariant $I$ of this black hole is of the form
\begin{eqnarray}
 I=\int \frac{dM}{\Delta\omega(E)}. \label{adiabatic}
\end{eqnarray}
Here, $\Delta\omega(E)$ still takes the form (\ref{vibrational}).
Substituting the thermodynamic quantities into (\ref{adiabatic}), we
obtain
\begin{eqnarray}
 I&=&\int \frac{dM}{2\pi T}\nonumber\\
  &\approx&\int 4M\bigg(1+\frac{4M^{3}}{\sqrt{\pi\theta^{3}}}
             e^{-\frac{M^{2}}{\theta}}\bigg)dM\nonumber\\
  &=&2M^{2}-\frac{8M^{2}}{\sqrt{\pi}}e^{-\frac{M^{2}}{\theta}}
     \left( \frac{M}{\sqrt{\theta} }
              +  \mathcal{O}\big(\frac{\sqrt{\theta}}{M}\big)
          \right)\nonumber\\
  &=&\frac{A}{8\pi} +  e^{-\frac{M^{2}}{\theta}}\mathcal{O}\big(\frac{\sqrt{\theta}}{M}\big).
\end{eqnarray}
With the Bohr-Sommerfeld quantization condition $I\approx
n\hbar\;\;(n\gg 1)$ \footnote{The quantum number $n$ is actually a
measure of the areal deviation from extremality for the black hole.
So, the black hole here is required to be far from its extremal
case, i.e., $M\gg M_{0}$.}, we get the area spectrum
\begin{eqnarray}
 A_{n}=8\pi\hbar \cdot n
     +e^{-\frac{M^{2}}{\theta}}\mathcal{O}\big(\frac{\sqrt{\theta}}{M}\big).
\end{eqnarray}
Recalling the entropy/area law, $S=\frac{A}{4}$, the quantized
entropy spectrum is obtained
\begin{eqnarray}
 S_{n}=2\pi \hbar\cdot n
     +e^{-\frac{M^{2}}{\theta}}\mathcal{O}\big(\frac{\sqrt{\theta}}{M}\big).
\end{eqnarray}
Obviously, the results are consistent with (\ref{areaspectrum}) and (\ref{entropyspectrum}),
which are obtained from the modified Hod's method.

Now, we have obtained the quantized area spectrum and entropy
spectrum for the noncommutative black hole with the modified Hod's
and Kunstatter's methods for the case $\frac{M}{\sqrt{\theta}}\gg 1$. Neglecting the high order of $\frac{\sqrt{\theta}}{M}$, both the
methods give the same equally spaced area spectrum and entropy
spectrum, which are in full agreement with that of the Schwarzschild black hole
given by Maggiore \cite{Maggioreprl2008}.

%%%%%%%%%%%%%%%%%%%%%%%%%%%%%%%%%%%%%%%%%%%%%%%%%%%%%%%%%%%%%%%%%%%%%%%%%%%%%%
\begin{figure*}
\centerline{\subfigure[]{\label{Hor}
\includegraphics[width=6.5cm,height=5cm]{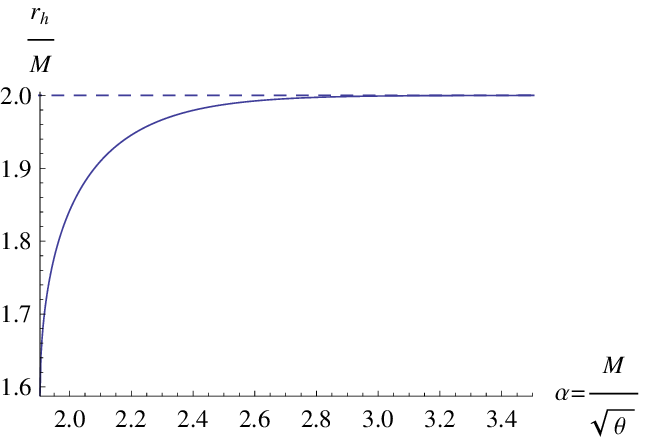}}
\subfigure[]{\label{Areasping}
\includegraphics[width=6.5cm,height=5cm]{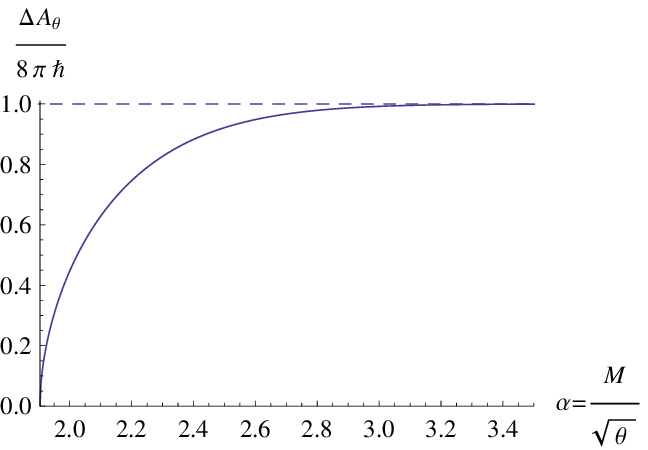}}}
\caption{(a) $\frac{r_{h}}{M}$ vs $\alpha=\frac{M}{\sqrt{\theta}}$ and (b) the area spacing vs $\alpha=\frac{M}{\sqrt{\theta}}$. The extremal black hole locates at $\alpha=1.90412$ and the minimum horizon $r_{h}=1.58732 M$.}
\end{figure*}
%%%%%%%%%%%%%%%%%%%%%%%%%%%%%%%%%%%%%%%%%%%%%%%%%%%%%%%%%%%%%%%%%%%%%%%%%%%%%%%%

Note that our results above are all for $\frac{M}{\sqrt{\theta}}\gg 1$. Here, we would like to give a brief discussion with the numerical method for small $\frac{M}{\sqrt{\theta}}$ limit. Note the range of $\theta$ (\ref{Range}), we define a new parameter $\alpha=\frac{M}{\sqrt{\theta}}\geq 1.90412$. The behavior of $r_{h}$ can be found in Fig. \ref{Hor}. From it, we can see that, the horizon $r_{h}<r_{\text{Sch}}=2 M$ and there exists a minimum horizon at $r_{h}=1.58732 M$, which corresponds to the extremal black hole. For the case $\frac{M}{\sqrt{\theta}}\gg 1$, the Schwarzschild black hole horizon will be recovered. On the other hand, with some calculations, we can express the area and entropy spectra in the forms
\begin{eqnarray}
 A_{n}&=&\Delta A_{\theta}\cdot n,\\
 S_{n}&=&\Delta S_{\theta}\cdot n.
\end{eqnarray}
We can see that the area and entropy spectra are discrete, however their spacings are dependent on the parameter $\theta$. The spacing of the area spectrum $\Delta A_{\theta}$ is described in Fig. \ref{Areasping}. It is monotonically increasing with $\frac{M}{\sqrt{\theta}}$. In the large $\frac{M}{\sqrt{\theta}}$ limit, the spacing of area spectrum $\Delta A_{\theta}=8\pi\hbar$, which is consistent with (\ref{areaspectrum}). While for the small $\frac{M}{\sqrt{\theta}}$, the area spectrum closely depends on $\frac{M}{\sqrt{\theta}}$. For $S=\frac{A}{4}$, the spacing of the entropy spectrum has the similar behavior as the spacing of the area spectrum.

\section{Summary}
\label{Discussion}

In this paper, we mainly deal with the noncommutative Schwarzschild
black hole spacetime, where the point-like structure is
eliminated and the point-like object is replaced by a smeared
object. This special property makes the noncommutative Schwarzschild
black hole behaves very different from the conventional one. We first
examine the thermodynamics quantities of the noncommutative black
hole. Its behaviors are very different from the conventional one
near the extremal case, while they meet each other far from the
extremal case. The quantization of the area and entropy for the noncommutative black
hole is also studied. We calculate the area spectrum and entropy
spectrum with the modified Hod's and the Kunstatter's
methods, respectively. The results show that (1) The area spectrum and entropy spectrum are discrete. (2) The spectrum spacings are dependent on the parameter $\frac{M}{\sqrt{\theta}}$. (3) The spacing of the area spectrum of the noncommutative Schwarzschild black hole is smaller than that of the conventional one. So does the spacing of the entropy spectrum. (4) The spectra from the two methods are the same. Especially, when $\frac{M}{\sqrt{\theta}}\gg 1$, the area and entropy spectra are consistent with that of the conventional Schwarzschild black hole. These results can help us to further understand the properties of the noncommutative black hole spacetime.

\section*{Acknowledgments}
%\acknowledgments

This work was supported by the Program for New Century Excellent
Talents in University, the National Natural Science Foundation of
China (No. 11075065), the Huo Ying-Dong Education Foundation of
Chinese Ministry of Education (No. 121106), the Doctoral Program Foundation of
Institutions of Higher Education of China (No.
20090211110028), and the Fundamental Research Funds for the
Central Universities (No. lzujbky-2009-54 and No. lzujbky-2009-
163).

\end{document}